\documentclass[a4paper, twocolumn]{revtex4-2}

\usepackage[utf8]{inputenc}
\usepackage{hyperref}
\usepackage{amsmath}
\usepackage{amssymb}
\usepackage{bm}
\usepackage{xcolor}
\usepackage{graphicx}
\usepackage{wrapfig}
\usepackage[acronym]{glossaries}
\usepackage{blindtext}
\usepackage{xfrac}
\usepackage{dsfont}

\usepackage{mathtools}

\DeclarePairedDelimiterX\braket[2]{\langle}{\rangle}{#1 \delimsize\vert #2}

\begin{document}
\title{Hyperfunction formulation of many body Green's functions and the Matsubara formalism}
\author{Roman Smit}
\affiliation{Institut f\"ur Theoretische Physik, Universit\"at Frankfurt, Max-von-Laue
	Strasse 1, 60438 Frankfurt, Germany}
\date{\today}

\begin{abstract}
	We show that the single-particle Green's functions used in many body theory have an elegant description in the form of hyperfunctions.
	We summarize the necessary hyperfunction concepts.
	We show that the analytical properties and the relations between different Green's functions are natural within this formulation.
	Important results from the standard formalism are recovered straightforwardly.
	We argue that hyperfunctions could possibly provide a powerful new tool for many body theory, if the formalism could be developed beyond single-particle Green's functions.
\end{abstract}

\maketitle

\ifx\forJMP\undefined 
\else 
\fi 

\section{Introduction}
The Matsubara formalism~\cite{Matsubara1955, fetter2003quantum, abrikosov2012methods} is a cornerstone of modern many body theory.
Quantum mechanical time evolution is Wick-rotated to imaginary times.
The time evolution operator then has the same form as a Boltzmann weight, which allows to write thermal expectation values as integrals over imaginary time.
For non-interacting systems, a Wick-theorem holds for imaginary time ordered expectation values, which ultimately enables diagrammatic perturbation theory at nonzero temperatures.

In the standard formalism, the Wick rotation is understood as an analytic continuation of the correlation functions to imaginary times.
Calculations are performed with imaginary times and frequencies.
Thereafter, the physically relevant correlations are obtained by analytic continuation back to real times and frequencies.
In the following, we want to present an alternative approach using hyperfunctions.
We concentrate on two-point correlation functions, which are ubiquitous in many body theory especially in the form of single-particle Green's functions.
The relations between the imaginary and real time and frequency domains are natural within the hyperfunction formulation.
The well known analytical properties and the relations between retarded, advanced, and Matsubara Green's functions become straightforward and the relation to physical observables is immediate.

The paper is structured as follows.
In Sec.~\ref{sec:hyperfunctions}, we informally introduce the concept of hyperfunctions.
In Sec.~\ref{sec:applied_hyperfunctions}, we present a more concrete formulation of hyperfunction theory developed in Ref.~\cite{imai2013applied}. This formulation is more practical for physicists, and we summarize some results that will be useful in the treatment of Green's functions.
In Sec.~\ref{sec:hyper_greens}, we show how the Green's function formalism can be described by hyperfunctions.
We work out the relations between the Matsubara, retarded and advanced Green's functions and the spectral density. Furthermore, we recover the important result that the retarded and advanced Green's functions in real frequency can be obtained by analytical continuation of the Matsubara Green's function from the imaginary Matsubara frequencies.
In Sec.~\ref{sec:outlook}, we discuss our findings and try to foreshadow how the hyperfunction approach to many body theory can be developed beyond single-particle Green's functions.

\section{Hyperfunctions}
\label{sec:hyperfunctions}
Hyperfunctions were introduced by Mikio Sato in 1958~\cite{Sato1959}, and an intuitive explanation due to Penrose can be found in Ref.~\cite{penrose2007road}.
In this section, we give an informal introduction to the basic concept.

\begin{figure}[h]
	\ifx\forJMP\undefined 
		\begin{minipage}{.4\linewidth}
			\includegraphics[width=\linewidth]{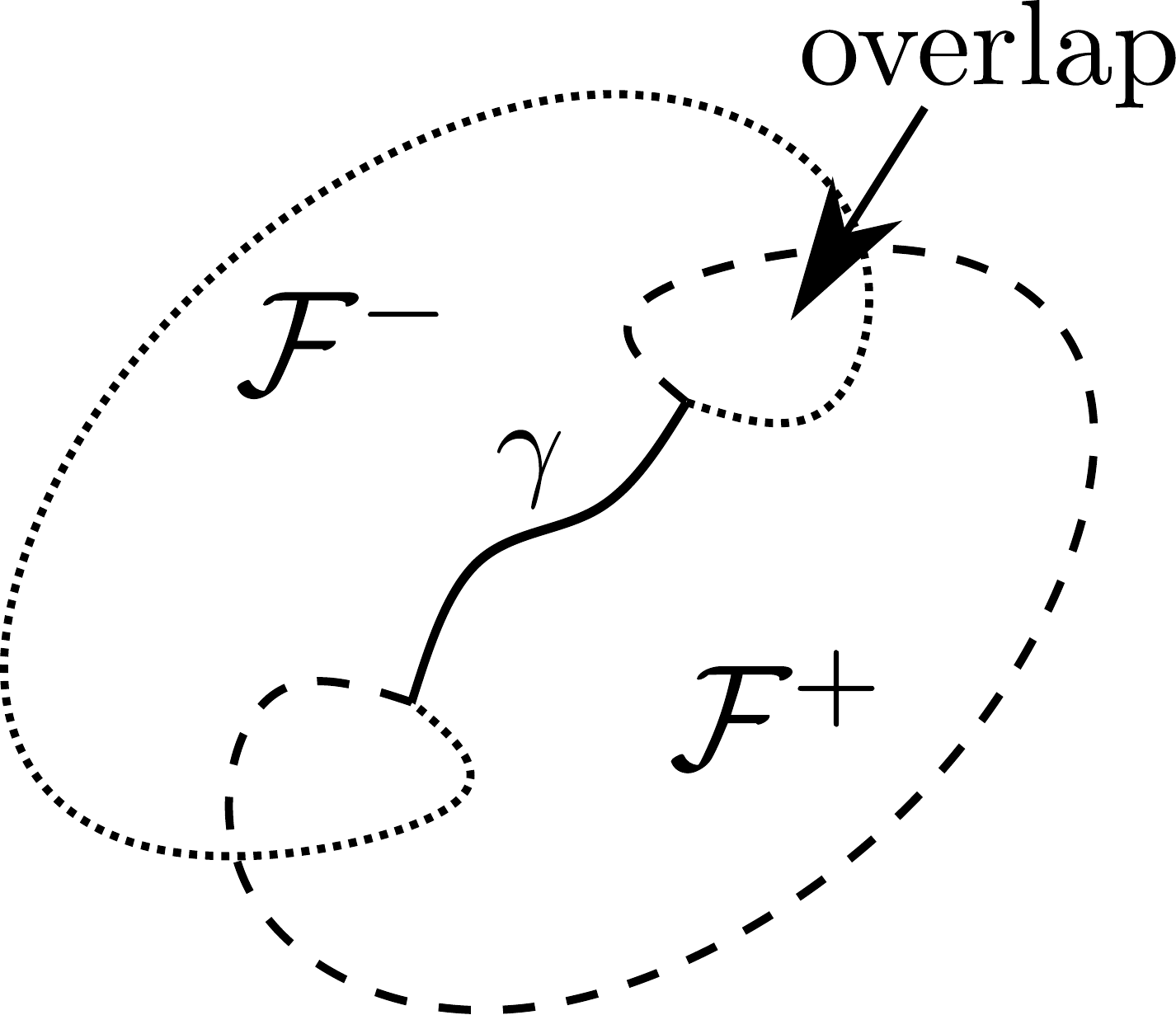}
		\end{minipage}
		\hspace{2mm}
		\begin{minipage}{.57\linewidth-2mm}
	\else 
		\begin{minipage}{.25\linewidth}
		\includegraphics[width=\linewidth]{fig_hyperfunction_regions}
		\end{minipage}
		\hspace{2mm}
		\begin{minipage}{.45\linewidth}
	\fi 
	\caption{Hyperfunctions are defined as equivalence classes of pairs of holomorphic functions on open regions $\mathcal{F}^+$, and $\mathcal{F}^-$ of the complex plane, which share a portion $\gamma$ of their boundaries.}
	\label{fig:hyperfunction_regions}
	\end{minipage}
\end{figure}

Consider two open regions $\mathcal{F}^+$ and $\mathcal{F}^-$ of the complex plane (or more generally of an arbitrary Riemann sheet) as depicted in Fig.~\ref{fig:hyperfunction_regions}. The regions share a portion $\gamma$ of their boundaries. Note that because both regions are open, neither of them includes $\gamma$. $\mathcal{F}^+$ and $\mathcal{F}^-$ may or may not overlap.
Now let $F^+$ be a holomorphic function on $\mathcal{F}^+$ and let $F^-$ be a holomorphic function on $\mathcal{F}^-$. A hyperfunction can then be defined as an equivalence class of the pair $(|F^+,F^-|)$. Two pairs of functions are equivalent if they differ only by a global function $\phi$ which is holomorphic on the union of $\mathcal{F^+}$, $\mathcal{F^-}$ and $\gamma$, i.e.
\begin{equation}
(|F^+,F^-|) \equiv (|F^+ + \phi,F^- + \phi|) \;.
\label{eq:def_equiv}
\end{equation}

Informally, a hyperfunction can be interpreted as the difference between $F^+$ and $F^-$ in the limit close to $\gamma$. Clearly, the difference does not change upon addition of a globally holomorphic function $\phi$.

Specifying $\gamma$ as the real line, and $\mathcal{F}^+$ and $\mathcal{F}^-$ as the upper and lower complex half plane respectively, one finds that any Schwartz-distribution $f$ of arbitrary order can be represented as~\cite{bremermann1965distributions}
\begin{equation}
\lim_{\epsilon \to 0^+} \int_{\infty}^{\infty} 
\left[F^+(x+i\epsilon) - F^-(x-i\epsilon)\right] \varphi(x) dx
= \langle f , \varphi \rangle \;,
\label{eq:schwarz_representation}
\end{equation}
where $\varphi$ is a test function in Schwartz's sense and $F^+$, $F^-$ are holomorphic on the upper and lower complex half planes respectively, excluding the real line.
The above relation shows that hyperfunctions offer an alternative approach to generalized functions, which are usually treated within distribution theory or as limits of sequence functions.
Indeed, by the above relation one can show that the Dirac delta $\delta(x)$ can be represented by $F^+(z) = F^-(z) = 1/2 \pi i z$, and the Heaviside step function can be represented by $F^+(z) = F^-(z) = -\log(-z)/2\pi i$. We remark that hyperfunctions are even more general than distributions, i.e. there are hyperfunctions which are not Schwartz distributions~\cite{bremermann1965distributions}.

\section{Applied hyperfunction theory}
\label{sec:applied_hyperfunctions}
Physical applications of hyperfunctions exist in relativistic quantum field theory~\cite{bremermann1965distributions}, fluid dynamics~\cite{imai2013applied}, and Twistor theory~\cite{Woit2021}.
Applications to statistical many body theory are to our knowledge not yet described in the literature.
To establish this connection, it will be crucial to understand the Fourier transform of a hyperfunction on the real line. Luckily, this was already worked out by Isao Imai~\cite{imai2013applied}. In this section, we summarize the terminology and relevant calculational tools as presented by Imai, with only slight adaptions.

First, we define the domains $\mathcal{F}^+$, $\mathcal{F}^-$ and $\gamma$. We choose $\gamma$ to be the real line or a section thereof.
$\mathcal{F}^+$ ($\mathcal{F}^-$) shall be an open set in the upper (lower) complex half plane whose boundary includes $\gamma$.
Then, we may simplify our notation. Because the domains $\mathcal{F}^+$ and $\mathcal{F}^-$ have no overlap, one can write
\begin{equation}
(|F^+(z), F^-(z)|) = F(z) \;,
\end{equation}
where $F(z)$ is a function which is holomorphic in both $\mathcal{F}^+$ and $\mathcal{F}^-$ but not necessarily on $\gamma$.

The connection to ordinary functions and distributions is established by virtue of Eq.~(\ref{eq:schwarz_representation}). We define the limit
\begin{equation}
f(x) = \lim_{\epsilon \to 0^+}
\left[F(x+i\epsilon) - F(x-i\epsilon)\right]
\;,
\label{eq:def_value}
\end{equation}
where $x$ is real. If this limit exists, we say that $f(x)$ is the value of the hyperfunction at $x$. Even if the limit does not exist, we denote the hyperfunction by $f(x)$ and call $F(z)$ the \textit{generating function}. We write
\begin{align}
f(x) = \text{HF}\, F(z) \;,\\
F(z) = \text{GF}\, f(x) \;,
\end{align}
where GF stands for generating function and HF stands for hyperfunction. Note that the generating function is not unique; different generating functions may generate the same hyperfunction due to the equivalence~(\ref{eq:def_equiv}).

The sum of two hyperfunctions $f(x)$ and $g(x)$ can straightforwardly be defined in terms of their generating functions $F(z)$ and $G(z)$ as
\begin{align}
f(x)+g(x)
&\equiv
\text{HF}\,  \left[ F(z) + G(z) \right]
\end{align}
It is obvious that if the value $f(x)+g(x)$ as defined by Eq.~(\ref{eq:def_value}) exists, it is the sum of $f(x)$ and $g(x)$, i.e. the hyperfunction sum reduces to the ordinary sum in the case of ordinary functions.
In a similar manner, one can define the multiplication of a hyperfunction $f(x)$ with an analytic function $\phi(x)$ by 
\begin{align}
\phi(x) f(x)
&\equiv \text{HF}\,  \phi(z) F(z)
\end{align}
where $\phi(z)$ is the analytic continuation of $\phi(x)$.
A general product between two hyperfunctions cannot be defined without restrictions to the hyperfunctions.

The derivative of a hyperfunction can be defined as
\begin{equation}
\partial_x f(x)
= \text{HF}\, \partial_z F(z)
\end{equation}
which again for ordinary functions reduces to ordinary differentiation.

\begin{figure}[h]
	\ifx\forJMP\undefined 
		\begin{minipage}{.55\linewidth}
		\includegraphics[width=\linewidth]{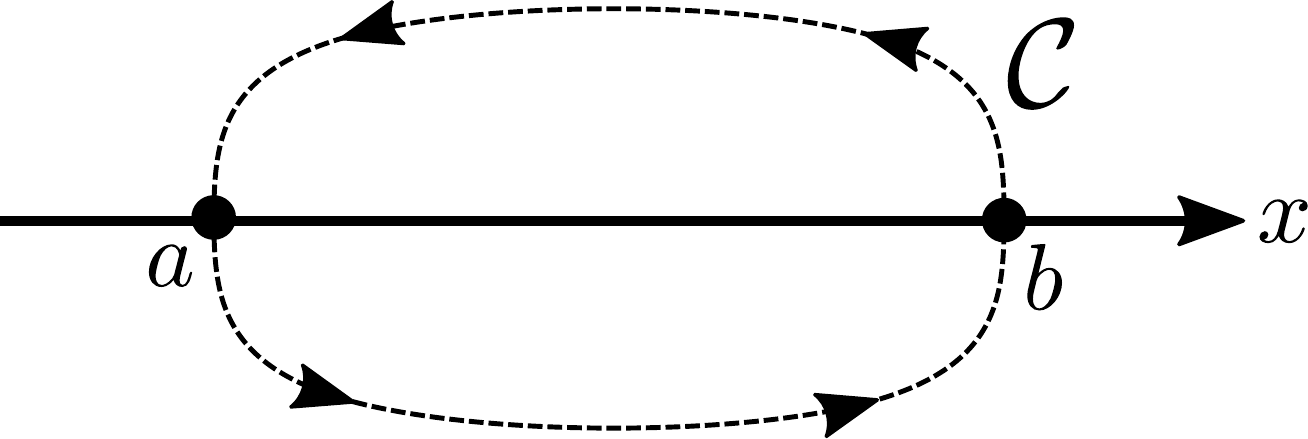}
		\end{minipage}
		\hspace{2mm}
		\begin{minipage}{.42\linewidth-2mm}
	\else 
		\begin{minipage}{.35\linewidth}
			\includegraphics[width=\linewidth]{fig_hyper_int_contour}
		\end{minipage}
		\hspace{2mm}
		\begin{minipage}{.35\linewidth}
	\fi 
	\caption{The definite integral of a hyperfunction is defined in terms of a closed contour integral of the generating function.}
	\label{fig:hyper_int_contour}
	\end{minipage}
\end{figure}

A definite integral over a hyperfunction $f(x)$ can be defined in terms of a contour integral of the generating function $F(z)$ as
\begin{equation}
\int_a^b f(x) dx \equiv -\int_\mathcal{C} F(z) dz \;,
\end{equation}
where the contour $\mathcal{C}$ is a loop enclosing the interval $[a,b]$ on the real line, as shown in Fig.~\ref{fig:hyper_int_contour}.
The equivalence to the integral of an ordinary function can be seen by deforming the contour towards the real line in the upper and lower half plane. Then, the reverse directions of the paths above and below the real line produces the minus sign in the definition of the value~(\ref{eq:def_value}) and the integral of an ordinary function is recovered.

\begin{figure}
	\ifx\forJMP\undefined 
		\begin{minipage}{.57\linewidth}
			\begin{tabular}{l}
				a) \\
				\includegraphics[width=\linewidth]{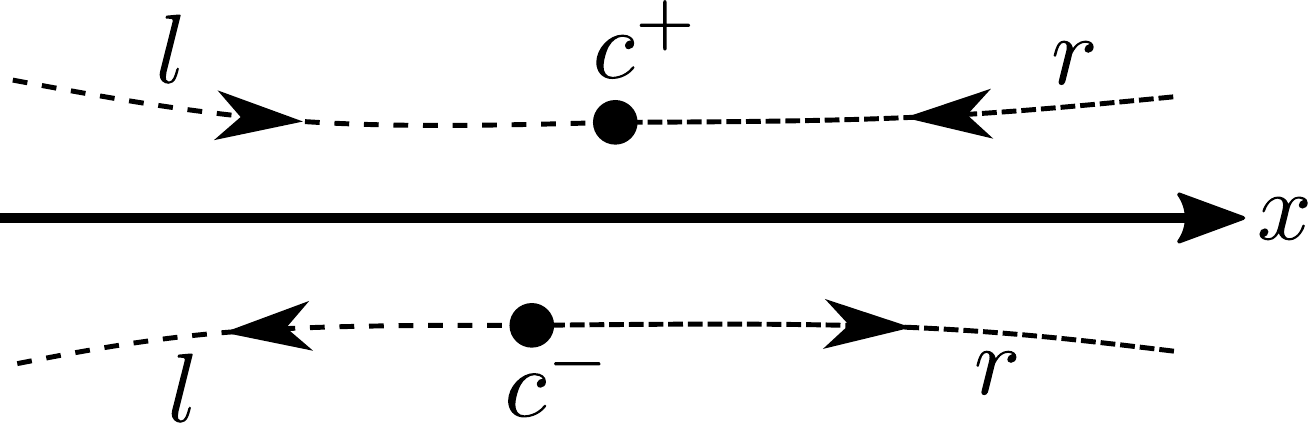} \\
				\vspace{1mm} \\
				b) \\
				\includegraphics[width=\linewidth]{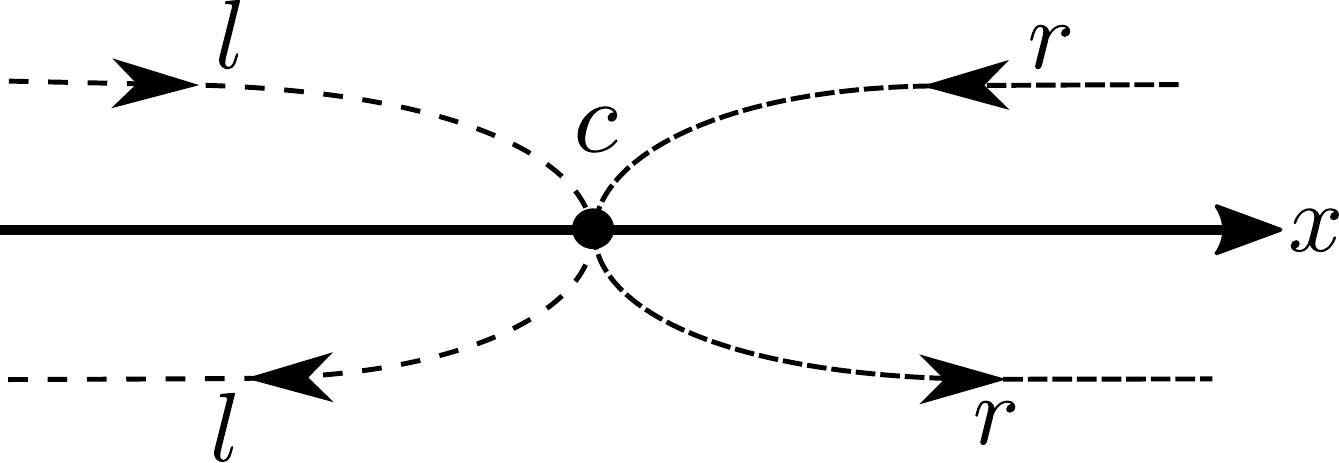}
			\end{tabular}
		\end{minipage}
		\hspace{2mm}
		\begin{minipage}{.4\linewidth-2mm}
	\else 
		\begin{minipage}{.4\linewidth}
			\begin{minipage}{6mm}
			a)
			\end{minipage}
			\begin{minipage}{\linewidth-7mm}
			\includegraphics[width=\linewidth]{fig_hyper_int_contour_infinity}
			\end{minipage}
			\\ \vspace{.5cm}
			\begin{minipage}{6mm}
				b)
			\end{minipage}
			\begin{minipage}{\linewidth-7mm}
				\includegraphics[width=\linewidth]{fig_hyper_int_contour_joined}
			\end{minipage}
			\begin{minipage}{1cm}
				\
			\end{minipage}
		\end{minipage}
		\hspace{2mm}
		\begin{minipage}{.3\linewidth}
	\fi 
	\caption{a) For the Fourier transform, the integration loop is extended to $\pm \infty$ and divided into left and right part at arbitrary points $c^+$ and $c^-$.
	b) Because $c^+$ and $c^-$ are arbitrary, they can be moved to join at a point $c$ on the real axis if the hyperfunction is regular there.}
	\label{fig:hyper_int_contour_fourier}
	\end{minipage}
\end{figure}

For an ordinary function $f(x)$, the Fourier transform is defined as
\begin{equation}
g(\xi) = \int_{-\infty}^{\infty} \frac{dx}{2\pi} f(x) e^{-i x \xi} \;.
\label{eq:ft}
\end{equation}
The Fourier transform can be generalized to hyperfunctions by defining the generating function $G(\zeta)$ as
\begin{subequations}
\begin{align}
G^+(\zeta) &= \int_l \frac{dz}{2\pi} F(z) e^{-i \zeta z} \;, \\
G^-(\zeta) &= \int_r \frac{dz}{2\pi} F(z) e^{-i \zeta z} \;,
\end{align}
\end{subequations}
where the integration contours $l$ and $r$ are as shown in Fig.~\ref{fig:hyper_int_contour_fourier}a.
The integration contours need some clarification. They result from extending the limits of a definite integral, which is a loop integral of the generating function, to $\pm \infty$. For $\text{Im}\,\zeta > 0$ the exponential of the Fourier transform converges only towards the left of the integration loop, while for $\text{Im}\,\zeta < 0$ it converges only towards the right.
Thus, the integration loop is split at arbitrary points points $c^+$ and $c^-$, dividing it into left and right part.
Defined this way, the hyperfunction $g(\xi)$ generated by $G(\zeta)$ is the Fourier transform of the hyperfunction $f(x)$ generated by $F(z)$. This definition reduces to the usual Fourier transform~(\ref{eq:ft}) in the case of ordinary functions.
While the generating function $G(\zeta)$ depends on the choice of $c^+$ and $c^-$, the hyperfunction $g(\xi)$ is independent of this choice. This can be proven by showing that a displacement of $c_+$ or $c_-$ simply adds an entirely holomorphic function to both $G^+(\zeta)$ and $G^-(\zeta)$, which does not change the hyperfunction.
As a consequence, the integration contour can be deformed such that $c_+$ and $c_-$ meet at a point $c$ on the real axis where $f(c)$ is regular, as shown in Fig.~\ref{fig:hyper_int_contour_fourier}b.

The inverse Fourier transform can be defined by
\begin{subequations}
	\label{eq:def_inv_hyper_FT}
\begin{align}
F^+(z) &= - \int_r d\zeta G(\zeta) e^{i \zeta z} \;, \\
F^-(z) &= - \int_l d\zeta G(\zeta) e^{i \zeta z} \;.
\end{align}
\end{subequations}
Note that due to the opposite sign in the exponential, the contours $l$ and $r$ are interchanged. Defined this way, the inverse Fourier transform of an ordinary function reduces to
\begin{equation}
f(x) = \int_{-\infty}^{\infty} d\xi \, g(\xi) e^{i x \xi} \;,
\end{equation}
which is the usual definition of the inverse Fourier transform.


\section{Hyperfunction formulation of Green's functions}
\label{sec:hyper_greens}
The central point of our discussion is a reinterpretation of the Matsubara Green's  function as the generating function of the spectral density. In the conventional formulation, the time $t$ is Wick rotated to a purely imaginary time $\tau=it$.
We will define the Matsubara Green's function in terms of the complex time variable
\begin{equation}
u = t - i\tau \;,
\end{equation}
where we include a negative sign in the imaginary part to be consistent with the usual Wick rotation.
Assuming a time independent hamiltonian, the Matsubara Green's function can be written as
\begin{equation}
G^M_{AB,\varepsilon}(u) = -\langle T_{\varepsilon} \left\{A(u)B(0)\right\} \rangle \;.
\end{equation}
Here, $\langle \cdot \rangle$ denotes a thermal expectation value. $A$ and $B$ are two time independent quantum operators, and $A(u)$ and $B(u)$ are their respective time evolutions in the Heisenberg picture, analytically continued to the complex time $u$. $T_{\varepsilon}$ is Wick's time ordering operator for imaginary times, which sorts larger $\tau$ to the left and accompanies each permutation of operators with a factor of $\varepsilon$. $\varepsilon$ can in principle be arbitrarily chosen as $+1$ or $-1$. However, in diagrammatic perturbation theory, where $A$ and $B$ are usually creation and annihilation operators, one has to choose $\varepsilon=+1$ for bosons and $\varepsilon=-1$ for fermions so that the Wick theorem can be used.

\begin{figure}[h]
	\ifx\forJMP\undefined 
		\begin{minipage}{.75\linewidth}
			\includegraphics[width=\linewidth]{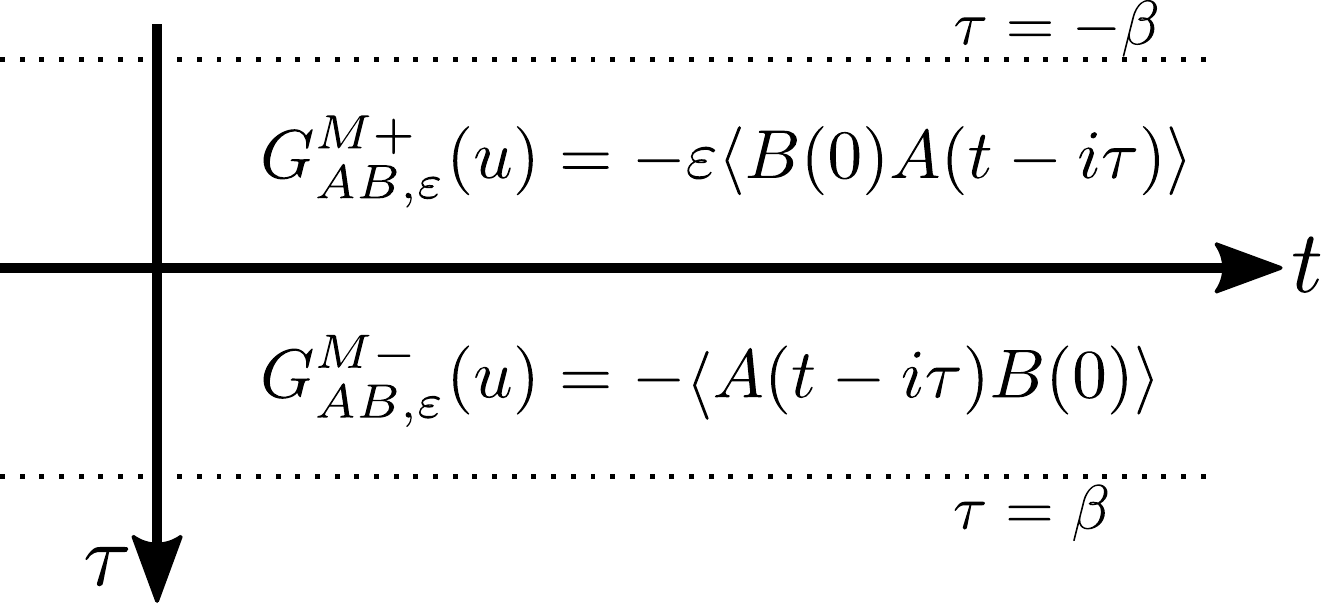}
		\end{minipage}
		\begin{minipage}{.75\linewidth}
	\else 
		\begin{minipage}{.4\linewidth}
			\includegraphics[width=\linewidth]{fig_hyper_matsubara_function}
		\end{minipage}
		\hspace{2mm}
		\begin{minipage}{.3\linewidth}
	\fi 
	\caption{Hyperfunction picture of the Matsubara Green's function, which is holomorphic in the intervals $-\beta < \tau < 0$ and $0 < \tau < \beta$.}
	\label{fig:hyper_matsubara_function}
	\end{minipage}
\end{figure}

The key observation is that $G^M_{AB,\varepsilon}(u)$ fulfills all properties of a generating function. To clarify this, we note that in the upper complex half plane of $u$ we have $\tau < 0$, while in the lower half plane we have $\tau > 0$. Thus by evaluating the time ordering in $G^M_{AB,\varepsilon}(u)$ explicitly, we may write
\begin{subequations}
\label{eq:def_GM}
\begin{align}
G^{M+}_{AB,\varepsilon}(u) &= -\varepsilon \langle B(0) A(u) \rangle \;, \\
G^{M-}_{AB,\varepsilon}(u) &= -\langle A(u) B(0) \rangle \;.
\end{align}
\end{subequations}
One can show via the spectral representations of the correlation functions that $G^{M+}_{AB,\varepsilon}(u)$ is holomorphic in the region $-\beta<\tau<0$ above the real line and $G^{M-}_{AB,\varepsilon}(u)$ is holomorphic in the region $0<\tau<\beta$ below the real line~\cite{mahan2000many, fetter2003quantum, nolting2009fundamentals}, where $\beta$ is the inverse temperature. We illustrated this hyperfunction structure in Fig.~\ref{fig:hyper_matsubara_function}.

We may now ask which hyperfunction is generated by $G^M_{AB,\varepsilon}(u)$. Calculating its values according to the definition~(\ref{eq:def_value}) leads to
\begin{align}
\lim_{\epsilon \to 0^+}
\left[G^{M}_{AB,\varepsilon}(t+i\epsilon) - G^{M}_{AB,\varepsilon}(t-i\epsilon)\right]
=& \langle \left[A(t), B(0) \right]_{-\varepsilon} \rangle
\nonumber \\
=& 2\pi \, S_{AB,\varepsilon}(t) \;,
\end{align}
where we denote by $[\cdot,\cdot]_{-}$ the commutator and by $[\cdot,\cdot]_{+}$ the anticommutator.
$S_{AB,\varepsilon}(t)$ is known as the spectral density, a central physical quantity which is directly related to spectroscopy experiments~\cite{nolting2009fundamentals}. 
We conclude that the Matsubara Green's function is a generating function of the spectral density,
\begin{equation}
2\pi \, S_{AB,\varepsilon}(t) = \text{HF} \, G^{M}_{AB,\varepsilon}(u) \;.
\end{equation}

\begin{figure}[h]
	\ifx\forJMP\undefined 
		\begin{minipage}{.75\linewidth}
			\includegraphics[width=\linewidth]{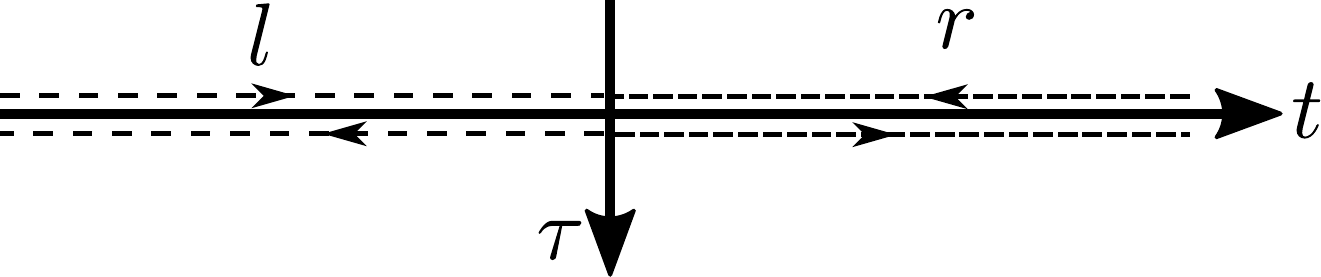}
		\end{minipage}
		\begin{minipage}{.75\linewidth}
	\else 
		\begin{minipage}{.4\linewidth}
			\includegraphics[width=\linewidth]{fig_hyper_int_contour_greens}
		\end{minipage}
		\hspace{2mm}
		\begin{minipage}{.3\linewidth}
	\fi 
	\caption{The integration contour of the Fourier transform is deformed such that it runs closely above and below the real axis.}
	\label{fig:hyper_int_contour_greens}
	\end{minipage}
\end{figure}

To reveal the analytical properties of the Green's functions, we perform the inverse hyperfunction Fourier transform from complex time $u$ to complex frequency
\begin{equation}
z = \omega + iy \;,
\end{equation}
and denote the transformed Matsubara Green's function by $\tilde{G}^M_{AB,\varepsilon}(z)$.
According to the definition~(\ref{eq:def_inv_hyper_FT}), $\tilde{G}^M_{AB,\varepsilon}(z)$ is given by
\begin{subequations}
	\label{eq:matsubara_inv_FT}
\begin{align}
\tilde{G}^{M+}_{AB,\varepsilon}(z) &= - \int_r du\, G^{M}_{AB,\varepsilon}(u) e^{i u z} \;, \\
\tilde{G}^{M-}_{AB,\varepsilon}(z) &= - \int_l du\, G^{M}_{AB,\varepsilon}(u) e^{i u z} \;,
\end{align}
\end{subequations}
where the integration contours $l$ and $r$ are as shown in Fig.~\ref{fig:hyper_int_contour_fourier}b.
We choose $c=0$ as the point where $l$ and $r$ cross the real axis and deform the contours towards the real axis as shown in Fig.~\ref{fig:hyper_int_contour_greens}.
\\
Inserting the definition of the Matsubara hyperfunction~(\ref{eq:def_GM}) and performing the limits $u \to t$ from above and below the real line, the integrals~(\ref{eq:matsubara_inv_FT}) may be written as
\begin{subequations}
\begin{align}
\tilde{G}^{M+}_{AB,\varepsilon}(z) =&
\int_{0}^{\infty} dt \left[
	- \varepsilon \langle B(0) A(t) \rangle + \langle A(t) B(0) \rangle
	\right] e^{itz} \;,
\\
\tilde{G}^{M-}_{AB,\varepsilon}(z) =&
\int_{-\infty}^{0} dt \left[
	\varepsilon \langle B(0) A(t) \rangle - \langle A(t) B(0) \rangle
	\right] e^{itz} \;.
\end{align}
\end{subequations}
In these expressions, we identify the retarded and advanced Green's functions
\begin{subequations}
\begin{align}
G_{AB,\varepsilon}^{\text{ret}}(t) &= - i \theta(t) \langle \left[A(t), B(0) \right]_{-\varepsilon} \rangle \;,
\\
G_{AB,\varepsilon}^{\text{av}}(t) &=  i \theta(-t) \langle \left[A(t), B(0) \right]_{-\varepsilon} \rangle \;,
\end{align}
\end{subequations}
which allows us to write
\begin{subequations}
	\begin{align}
	\tilde{G}^{M+}_{AB,\varepsilon}(z)
	&= i\int_{-\infty}^{\infty} dt\,
	G_{AB,\varepsilon}^{\text{ret}}(t)
	e^{itz} \;,
	\\
	\tilde{G}^{M-}_{AB,\varepsilon}(z)
	&= i\int_{-\infty}^{\infty} dt\,
	G_{AB,\varepsilon}^{\text{av}}(t)
	e^{itz} \;.
	\end{align}
\end{subequations}
This reveals an exact analogy with the standard formalism, where the retarded and advanced Green's functions in frequency space can be written in terms of one unified Green's function $G_{AB,\varepsilon}(\omega + iy)$, such that in the upper complex half plane $G=G^{\text{ret}}$ and in the lower half plane $G=G^{\text{av}}$, with a branch cut on the real line. Hence in the complex frequency domain, we identify the hyperfunction Fourier transform of the Matsubara Green's function with the Green's function,
\begin{equation}
\tilde{G}^{M}_{AB,\varepsilon}(z) = i G_{AB,\varepsilon}(z) \;,
\label{eq:ident_G_omega}
\end{equation}
up to a factor of $i$.

Per construction, $\tilde{G}^{M}_{AB,\varepsilon}(z)$ is a generating function for the spectral density in the frequency domain.
By virtue of the identification~(\ref{eq:ident_G_omega}), this is equivalent to
\begin{align}
2\pi S_{AB,\varepsilon}(\omega)
&=
\lim_{\epsilon \to 0^+} i\left[ G_{AB,\varepsilon}(\omega + i\epsilon) - G_{AB,\varepsilon}(\omega - i\epsilon)\right]
\nonumber \\
&= i G^{\text{ret}}_{AB,\varepsilon}(\omega) - i G^{\text{av}}_{AB,\varepsilon}(\omega) \;,
\end{align}
which is an important result known from the standard formalism. In the case of a real valued spectral density, this directly leads to the Kramers-Kronig relations.

\begin{figure}
	\ifx\forJMP\undefined 
		\begin{minipage}{.75\linewidth}
			\includegraphics[width=\linewidth]{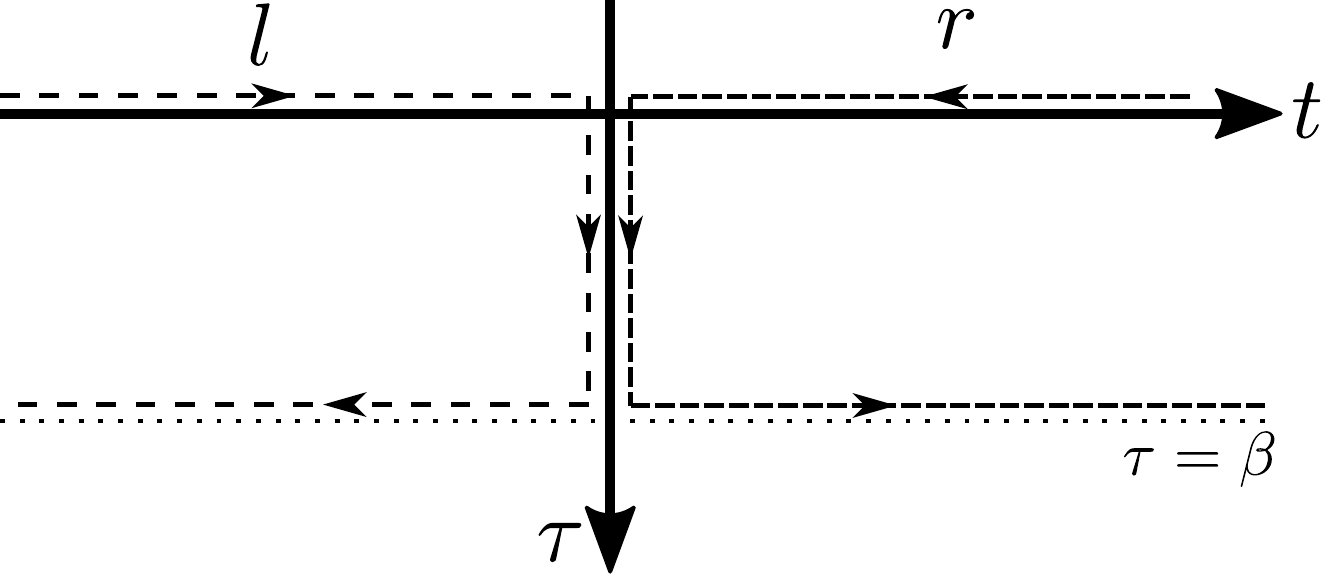}
		\end{minipage}
		\begin{minipage}{.75\linewidth}
	\else 
		\begin{minipage}{.4\linewidth}
			\includegraphics[width=\linewidth]{fig_hyper_int_contour_matsubara}
		\end{minipage}
		\hspace{2mm}
		\begin{minipage}{.3\linewidth}
	\fi 
	\caption{The integration contour of the Fourier transform is deformed towards the real axis in the upper complex half plane, and towards the line $\tau = \beta$ in the lower half plane.}
	\label{fig:hyper_int_contour_matsubara}
	\end{minipage}
\end{figure}

To conclude this section, we want to show the connection of the hyperfunction Fourier transform to the Matsubara frequencies.
Usually, the Matsubara Green's  function in frequency space is obtained as a Fourier series of the imaginary time Matsubara Green's  function on the interval $-\beta < \tau < \beta$, which leads to the discrete Matsubara frequencies\cite{Mermin1961, mahan2000many, nolting2009fundamentals}.
We therefore evaluate the inverse hyperfunction Fourier transform of the complex time Matsubara Green's  function~(\ref{eq:matsubara_inv_FT}) at the imaginary Matsubara frequencies $z = i\omega_n$. Here, $\omega_n$ is a bosonic Matsubara frequency if $\varepsilon=+1$ and a fermionic Matsubara frequency if $\varepsilon=-1$, i.e.
\begin{equation}
\omega_n = 
\begin{cases}
\frac{2n \pi}{\beta} & \text{if } \varepsilon=+1 \\
\frac{(2n+1)\pi}{\beta} & \text{if } \varepsilon=-1
\end{cases}
\;,
\end{equation}
with arbitrary integer $n$.
For the exponential in the Fourier transform, we then get a relation reminiscent of the Kubo-Martin-Schwinger condition\cite{Kubo1957, Martin1959, Haag1967},
\begin{equation}
e^{i(t-i\tau)(i\omega_n)} = \varepsilon e^{i(t-i\tau + i\beta)(i\omega_n)} \;.
\end{equation}
Recalling that the Matsubara Green's  function also fulfills the Kubo-Martin-Schwinger condition
\begin{equation}
\tilde{G}^{M}_{AB,\varepsilon}(t-i\tau) = \varepsilon \tilde{G}^{M}_{AB,\varepsilon}(t - i\tau + i\beta)
\end{equation}
for $0<\tau<\beta$, we see that the integrand of the Fourier transform is invariant under a complex time shift of $u \to u + i\beta$.
We leverage this invariance by deforming the integration contours $l$ and $r$ in the lower half plane towards the line $\tau = \beta$,  as shown in Fig.~\ref{fig:hyper_int_contour_matsubara}.
Then, the contour integrals near the real line cancel with the integrals near $\tau=\beta$ and only the integrals along the imaginary axis from $0$ to $-i\beta$ are left.
The hyperfunction Fourier transform~(\ref{eq:matsubara_inv_FT}) thus reduces to
\begin{equation}
\tilde{G}^{M\pm}_{AB,\varepsilon}(i\omega_n)
=
i \int_0^{\beta} d\tau\, G^{M}_{AB,\varepsilon}(-i\tau) e^{i \tau \omega_n} \;,
\label{eq:hyper_matsubara_fourier}
\end{equation}
where the upper (+) and lower (-) branch of $\tilde{G}^{M\pm}_{AB,\varepsilon}(i\omega_n)$ can be written in the same expression since the integration contours are the same. There is a subtlety with the bosonic zero frequency, as it lies neither in the upper nor in the lower half plane. However, the limit of $i\omega_n \to 0$ is the same from above and below.
Eq.~(\ref{eq:hyper_matsubara_fourier}) is identical to the Fourier series of the imaginary time Matsubara Green's  function known from the standard formalism up to a factor of $i$.
The factor of $i$ is the same as encountered before in Eq.~(\ref{eq:ident_G_omega}),
which proves that the Fourier components of the Wick rotated Matsubara Green's  function are given by the values of the complex frequency Green's function at the Matsubara frequencies.
This identification is familiar from the standard formalism and implies that the retarded and advanced Green's functions in real frequency can be obtained by analytical continuation of the Matsubara Green's function from the imaginary Matsubara frequencies to the real axis.

\section{Discussion and Outlook}
\label{sec:outlook}
We have shown that single-particle Green's functions can be described elegantly by hyperfunctions.
The spectral density can be interpreted as a hyperfunction which is generated by the Matsubara Green's function.
The analytical structure of Matsubara, advanced, and retarded Green's functions in time and frequency, as well as their relations to each other can be understood as natural consequences of the hyperfunction Fourier transform.
Within the hyperfunction formulation, it is straightforward to show that advanced and retarded Green's functions in real frequency can be obtained by analytic continuation from the imaginary Matsubara frequencies.

Up to the level of our analysis, the hyperfunction formalism represents a surprising simplification of the conventional formalism.
The question is whether the hyperfunction formulation can be developed beyond single-particle Green's functions.
Since hyperfunctions can be defined in more than one complex dimension~\cite{bremermann1965distributions}, it seems likely that higher order Green's functions might also have an elegant hyperfunction description.
Consequently, this may lead to a hyperfunction description of diagrammatic perturbation theory and related methods, or even to novel approximations.
Another possibility could be a hyperfunction formulation of non-equilibrium correlations, where the Keldysh-formalism\cite{Keldysh1964ud} could serve as a starting point.

It is hard to predict where a further development of the hyperfunction approach might lead.
In any case, the hyperfunction formulation of Green's functions seems to be a promising contender for a conceptually new approach to many body theory.

\section*{Acknowledgments}
I want to thank Peter Kopietz, Raphael Goll and Jessica Friedl for valuable discussions and suggestions concerning the form of this paper.

\ifx\forJMP\undefined 
\else 
\section*{Author Declarations}
The author has no conflicts to disclose.
\\
\fi 

\bibliographystyle{ieeetr}

\bibliography{bibliography}

\end{document}